\documentclass[aps,epsfig,prd,nofootinbib]{revtex4}

\fussy
\def\section{\@startsection {section}{1}{\z@}{-3.5ex plus -1ex minus 
    -.2ex}{2.3ex plus .2ex}{\bf }}
\def\subsection{\@startsection{subsection}{2}{\z@}{-3.25ex plus -1ex minus 
   -.2ex}{1.5ex plus .2ex}{\it }}

\usepackage{graphicx}
\newcommand{\CP }{ $\mathcal{CP}\ $}
\newcommand{\bpk}{$B\to J/\psi K_S \ $}
\newcommand{\gmt}{$g-2\ $}

\def \beq{\begin{equation}}         \def \eeq{\end{equation}}
\def \be{\begin{eqnarray}}          \def \ee{\end{eqnarray}}
\def \beqa{\begin{eqnarray}}     \def \eeqa{\end{eqnarray}}

\def \bea{\begin{array}}        \def \eea{\end{array}}

\def\a{\alpha} 
\def\b{\beta}  
\def\g{\gamma} 
 
\def\m{\mu}    
\def\n{\nu}    
\def\p{\pi}

\def \l{\lambda}

\def \abs#1{\left| #1 \right|}

\def\prd#1#2#3{    { Phys. Rev. }{\bf D#1}, #3 (#2)}

\def\prl#1#2#3{    { Phys. Rev. Lett. }{\bf #1} ,#3 (#2)}

\begin{document}
\title{Implications from recent measurements on $\sin 2\beta$ and muon $g-2$}

\author{Yue-Liang Wu}
\author{Yu-Feng Zhou\footnote{ Talk delivered by Y.F.
Zhou at International Conference on Flavor Physics
(ICFP2001),  May 31-June
6, at  Zhang-Jia-Jie, China. }}
\address{ Institute of Theoretical Physics, Chinese Academy of
Science, Beijing 100080, China }

\maketitle

\begin{center}
\begin{minipage}{4in}{\small
The recent data on \CP asymmetry in \bpk and muon \gmt are discussed in
the framework of standard model and beyond. Possible new phase effects
besides the CKM phase are discussed in the processes concerning \CP
violation in B decays and muon anomalous magnetic moment (muon \gmt).
It is found that the new phases will result in difference between
angles $\b$ measured from \bpk and the one from global fit. The ration
between them serves as a probe of not only new physics but also new
phase besides the CKM phase In the case of muon \gmt, the new phase
may change the interferences between various contributions. By
including the new phases, some cancelations in the real coupling cases
can be avoided and the large value of \gmt observed in the recent
measurements can be understood.}
\end{minipage}
\end{center}
\vskip 0.5cm

In the recent months, exciting experimental results were
obtained\cite{aubert:2001nu,abe:2001xe} in both heavy and light
flavour sectors \cite{brown:2001mg}. With the successfully running of B
factories, the time dependent CP asymmetries of decay mode \bpk has
been observed. It is the first time that CP violation is observed
beyond the Kaon system. The result although coincides with the Standard
Model (SM) predictions, implies the possibility of new physics beyond
the SM.
%
It is known that the SM has achieved great success in
phenomenology. However it has not yet been thoroughly tested. Up to
now, we still know very little about the intrinsic physics of
electro-weak symmetry breaking, the hierarchy of quark lepton mass
as  well as  the origin of \CP violation.   In this talk we will
discussed the possible new physics contributions to \bpk and muon
\gmt. we will focus on new phases beyond the CKM phase in the new
physics models.


In the SM,  by using the Wolfenstein parameterization
\cite{Wolfenstein:1983yz} the CKM matrix can be described by
four parameters $\l$, $A$, $\rho$ and $\eta$.
 As it is unitary, the elements of the matrix must satisfy the unitarity
 relations.
For example, for the columns containing $b$ quarks the
 unitarity relation is given by
$
V_{ud}V^*_{ub}+V_{cd}V^*_{cb}+V_{td}V^*_{tb}=0.
$

In the complex plane the relation  can be depicted as a closed
triangle
which  is often referred as Unitarity Triangle (UT) and the three
angles are defined as $\a$, $\b$ and $\g$ respectively.  
It may be interesting to measure
those angle independently and check if their sum is equal to $\p$.
That is the simplest way to test the SM.  However, the measurement
is not easy to be made. There are large difficulties from both
theoretical and experimental sides. It is not easy to find a
channel which has large branching ratio and small theoretical
errors. Fortunately, people found one decay mode of $B\to J/\psi
K_S$ which is  suitable for measuring the angle $\b$.  In this
channel, the branching ratio can reach $10^{-3}$ and the
theoretical error is less than a few percent.




The angle $\b$ can be  measured through
  time dependent CP asymmetry ${\cal A}_{CP}$  in decay mode
$B\to J/\psi K_S$.
The major goal of the current running $B$ factories is to precisely determine the
value of $\sin 2\b$.  The First preliminary results in the year 2000 from
BaBar and Belle \cite{Hitlin:2000tm,abashian:2001pa} were  quite smaller
than peoples expectation, the BaBar results was  almost compatible with zero.
However, the situation changed quickly in the beginning of the  year 2001.
 The improved data gave the average  value of $\sin 2\b \sim 0.47$.
  A few months ago , both groups reported their latest  results.
  This time the Belle claimed the observation
of large $\sin 2\b \sim 0.99$\cite{abe:2001xe}, which seemed  not  supported by the BaBar Collaboration. The
BaBar result is near $0.59$\cite{aubert:2001nu}.   
Now, the latest world average value of $\sin 2\b$ from time dependent \CP violation
in decay $B\to J/\psi K_S$ is
$ \sin 2\beta_{J/\psi}= 0.79\pm 0.1.   $

 In the SM, the angle $\beta^{\rm{SM}}_{\rm{fit}}$ extracted from the global
fit should be the same as the one measured from the time dependent
CP asymmetry in the decay $B\to J/\psi K_S$.  However, if there
exists new physics beyond the SM, the situation may be quite
different, the angle ``$\beta$'' extracted from two different
approaches are in general not equal. This is because $\epsilon_K$
and $\Delta m_B$ as well as $B\to J/\psi K_S$  will receive
contributions from new physics in a quite different manner
comparing to the ones from the SM. As a consequence, the extracted
``$\beta$'' from two ways will become different. In the most
general case, one may find that $\beta^{\rm{SM}}_{\rm{fit}} \neq
\beta_{J/\psi}\neq \beta$ when new physics exists and the ratio
\be
R_{\beta}\equiv{\sin 2\beta_{J/\psi} \over \sin
  2\beta^{\rm{SM}}_{\rm{fit}}}
\ee
is therefore not equal to unity.  

In a recent work\cite{wu:2001qu}, it
is found that the ratio $R_{\beta}$ may provide a useful tool for
probing new physics. For a detailed consideration, two interesting and
typical models are investigated. The one is the model with minimal
flavor violation (MFV)
~\cite{buras:2000xq,buras:2001af,buras:2001pn,Ali:2001ej} which
corresponds to the models without new phase beyond the SM such as
minimal SUSY models. Another one is the simple extension of SM with
two Higgs doublets (S2HDM) motivated from spontaneous
\CP violation ~\cite{wolfenstein:1994jw,wu:1994ja,wu:1999fe}.
In the models with MFV,  the value for $R_\b$ can be directly calculated and
has the following simple form\cite{wu:2001qu}
$
\left. R_{\beta}\right|_{MFV}\simeq{1- 0.79 \ \bar{\eta}\over
 1- 0.79 \ \bar{\eta}^{\rm{SM}}_{\rm{fit}} }
$
In this kind of model, as there is no additional \CP violating phase,
the $\sin 2\b_{J/\psi} $ which is measured from $B\to J/\psi K_{S}$
should not be affected and equal to the true value of $\sin
2\b$. Provided that the semileptonic $B$ decay $B\to X_u (\p,\rho)
\ell \n_\ell \ \ (\ell=e,\m )$ is not likely to be affected by new physics,
then the true value of $\bar{\eta}$ can be obtained. It is interesting to point
out that the deviation of $R_\b$ from unity from such model is alway
in the opposite sign with the measure $R_\b$. i.e. if the measure
$R_\b$ is lower than $1$, i.e., $R_\b < 1$  the models with MFV will predict a value
 larger than $1$, i.e., $R_{\beta}|_{MFV} > 1$, vice versa.
This simple observation indicates that
if the experiment obtains a value of $R_\beta$ different from unity, it
will be not only a signal of new physics but also a signal of new
physics beyond the MFV.  In the models with rich sources of \CP
violation, such as the two-Higgs-doublet model with spontaneous \CP
violation (S2HDM), the possibility of $R_\b \neq 1$  can be easily accommodated.

Recently, the E821 Collaboration at BNL has reported their
improved result on the measurement of muon anomalous magnetic
moment $a_\mu=(g_\mu-2)/2$ \cite{brown:2001mg}.  The difference between
the measurement and the Standard Model (SM) prediction
is reported \
\beq
\Delta a_\mu=a^{exp}_\mu-a^{SM}_\mu
=426\pm165 \times 10^{-11},
 \eeq
 which shows a relative large deviation ($\sim$2.6 $\sigma$) from the SM calculation. 
 As muon is about 210 times heavier than electron, it is expected that the
new physics effects on muon anomalous moment may be considerable.
A large amount of works have been made on checking the new physics
contributions to $\Delta a_\mu$ from various models.   Here we shall take  the standard model with two Higgs
doublets (S2HDM) motivated from the study of origin and mechanism
of CP violation\cite{wolfenstein:1994jw,wu:1994ja} as an example to discuss the new phase effects.  In this kind of model, the
Higgs sector of SM is simply extended by including an additional
Higgs doublet with all the Yukawa couplings being real. After
spontaneous symmetry breaking, the origin of both fermion masses
and CP-violating phases can be attributed to the well known Higgs
mechanism with a single CP-phase between two vacuum expectation
values. Namely CP symmetry is broken spontaneously \cite{Lee:1973iz,Lee:1974jb}.
This leads to the fact that  all the effective Yukawa couplings in the physical basis
become complex.  
 
Contributions to muon $g-2$ from the 2HDM of type II have been
discussed in Refs.\cite{Dedes:2001nx,dedes:2001hh}.  It seems that the one loop
contribution is not large enough to explain the data. Even for a
large value of $\tan\beta\sim 50$, one still needs a very light
mass of the scalar $M_h\sim 5 $ GeV.  If both the scalar and
pseudo-scalar are included, due to the cancelation between them,
the situation will be even worse.  One then needs to consider two
loop contributions. It was found  \cite{chang:2000ii,cheung:2001hz} that from the
two-loop diagrams of Barr-Zee type\cite{BZ}, the contributions
from pseudo-scalar is positive and could be larger than its
negative one loop contributions. Thus the net effects for
pseudo-scalar exchange become positive. Provided a sufficient
large value of the coupling $\tan\beta$, its contributions can
reach the experimental bound. By including the two loop diagrams,
the pseudo-scalar mass must be below $80$ GeV even when
$\tan\beta$ is large and around $50$ \cite{cheung:2001hz}. To avoid the
cancelation between scalar and pseudo-scalar exchange, the mass
of the scalar boson has to be pushed to be very heavy ( typically
greater than 500 GeV).

  In the S2HDM, as
there are flavor changing scalar interactions for both neutral and
charged scalar bosons, the bounds on the neutral Higgs mass can be
released through the inclusion of the internal $\tau$ loop at one loop
level \cite{tau}.  Besides this, as it will be shown below that the
complex and flavor dependent Yukawa couplings in S2HDM may completely
change the interference between one and two loop diagram
contributions.  As a consequence, one and two loop contributions can
be both positive and provide significant contributions to $\Delta
a_\mu$.



The phase effects of the Yukawa coupling has been discussed
in the literatures
\cite{wolfenstein:1994jw,wu:1994ja,wu:1999fe,Bowser-Chao:1998yp,wu:1998ng}.
The complex couplings may lead to sizable CP violation in charged and
neutral Higgs boson exchanging processes in quark sector, such as
$b\to s\gamma$\cite{wolfenstein:1994jw,wu:1998ng,bsg},  neutral meson mixings\cite{wu:1999fe}, etc. In
the lepton sector, it may result in large electric dipole moment of
leptons \cite{Bowser-Chao:1998yp,ylw,iltan:2001nk}.


%
In the case of effective real Yukawa couplings for neutral scalar
bosons, $h$ is purely a scalar while $A$ is a pseudo-scalar. The
contributions to muon anomalous moment from these two particles
always have different signs at both one loop and two loop level.
Namely, for scalar $h$-exchange, the contribution is positive from
one loop and negative from two loop. But for pseudo-scalar
$A$-exchange , the one loop contribution is always negative and
the two loop contribution is positive. As the experimental data
indicate that the deviation of $g-2$ from SM is positive.  The
large two loop contribution may lead to the conclusion that the
$A$-exchange must be dominant and the $h$-exchange must be made to
be negligible small.

However, in the general case, the situation can be quite
different. As all the couplings are complex in S2HDM, the Yukawa
couplings for neutral scalar bosons $h$ and $A$ contain both
scalar and pseudo-scalar type fermion interactions.  Further more,
the couplings are all flavor dependent, which is far away from the
2HDM of type I and I$\!$I, where the couplings are given by a
simple function of $\tan\beta$. The fact that one loop and two
loop contributions depend on different complex Yukawa couplings
provides an alternative possibility that they may not always
cancel each other. In a large parameter space, the two kind of
contributions may be constructive and result in a large value of
muon $g-2$

 The one loop flavor conserving contributions from the scalar and pseudo-scalar
 have been investigated in Refs. \cite{oneloop}.
In S2HDM, the results for $h$ are given by
\beqa
 \Delta a^h_\mu&=&{g^2\over 32 \pi^2} {m^4_\mu \over m^2_W m^2_h}
        \left[ (\mbox{Re}\xi_\mu)^2 \left( \ln {m^2_h\over m^2_\mu}-{7\over6}\right)
                      -(\mbox{Im} \xi_\mu)^2 \left( \ln {m^2_h\over m^2_\mu}-{11\over6}\right)
                \right] 
\eeqa
where $g$ is the weak gauge coupling constant.   
It can be seen that its contributions
decrease rapidly with $m_h$ increasing, which means that the one
loop diagram cannot give large contributions to $\Delta a_\mu$
except for a very small value of $m_h\sim $  10 GeV or a very
large value of $|\xi_\mu|\sim $ 70. Note that in the S2HDM, the
contributions from $h$ and $A$  can be negative and positive
depending on the sign of $\cos 2\delta_\mu$, which is completely
different from the other type of 2HDM, such as type I and type
I$\!$I, where the $h(A)$ loop diagrams always give positive
(negative) contributions.

In the case of one loop flavor changing processes, the loops with
internal $\tau$-lepton play an important role as it is much heavier
than the $\mu$-lepton.  This then leads to an enhancement factor of
the order $m_\tau^2/m^2_\mu\sim {\cal O}(10^2)$ relative to the flavor
conserving one.  In S2HDM, the contributions from $h(A)$ exchange are
given by
\beqa
\Delta a^{h(A)}_\mu&=&\pm{g^2 \over 32 \pi^2}
{m^2_\mu m^2_\tau  \over m^2_W m^2_{h(A)}}
     \left(\ln {m^2_{h(A)}\over m^2_\tau}-{3\over2}\right)
     \abs{\xi_{\mu\tau}}^2 \cos 2\delta_{\mu\tau}
\eeqa
 In obtaining the above expression, we have taken
$\xi_{\mu\tau}=\xi_{\tau\mu}$ for simplicity.  The contributions to  $\Delta a_\mu$ may be
considerable large when $|\xi_{\mu\tau}|$ is large. For
$|\xi_{\mu\tau}|= 30\sim 50$,  the recently reported 2.6 sigma
experiment vs. theory deviation can be easily explained even for a
heavy scalar boson $m_h > 100$ GeV. So far, there are no strict
constraints on the values of the coupling $|\xi_{\mu\tau}|$.
Studies on the rare decays $\tau\to \mu(e) \gamma$ and  $\tau \to
3\mu(e)$ and the electric dipole moment of $\tau$ will be useful
to provide an interesting constraint on the parameter. However, as
the relevant experimental data at present are primitive and such
processes often contain more couplings such as $\xi_\tau$ and
$\xi_{\tau e }$ , the resulting constraints can not be clearly
obtained and they are not yet be very strong. To obtain the upper
bounds on $\xi_{\tau\mu}$, further studies are needed.

For the case that the fermion
loop is dominated by top quark, the Barr-Zee type two loop diagrams
may lead to a sizable $\Delta a_\mu$. In S2HDM, their contributions
are given by 
\beqa \Delta a^h_\mu &=& { N_c q^2_t \alpha^2 \over
4\pi^2 \sin^2 \theta_W} {m^2_\mu \over m^2_W} \left[ \mbox{Im}\xi_t
\mbox{Im}\xi_\mu f\left( {m^2_t\over m^2_h }\right) -\mbox{Re}\xi_t
\mbox{Re}\xi_\mu g\left( {m^2_t\over m^2_h }\right) \right]
\eeqa
 where $N_c=3$ is the color number and
$q_t=2/3$ is the charge of top quark.  $\alpha=1/137$ and
$\sin^2\theta_W=0.23$ are the fine structure constant and weak
mixing angle respectively. $f(z)$ and $g(z)$ are two integral
functions which  can be found in Ref. \cite{BZ}  
  Unlike in the one loop case, where $\Delta
a_\mu$ decreases rapidly with growing $m_h$, the two loop
contributions decrease relative slowly and their signs depend on
the value of $\delta_{\mu\tau}$.  Therefore for a very large value
of $m_h$, the two loop effects become dominant. Another difference
is that in the one loop case the new physics contributions only
depend on $\xi_\mu$, while the two loop contributions depend on
two couplings $\xi_\mu$ and $\xi_t$, if the fermion loops for $b$
or $\tau$ are included \cite{chang:2000ii,cheung:2001hz}, it will depend on more
parameters.  As the two couplings are complex numbers, the
interference between one and two loop diagrams may not always be
destructive. There exists a large parameter space in which the one
and two loop contributions are all positive. Thus they can result
in a large contribution to $\Delta a_\mu$ .  The constraint of
$|\xi_t|$ has been studied in $B^0-\bar{B}^0$ mixing and $b\to s
\gamma$ as well as the neutron electric dipole moment
\cite{wu:1998ng,wolfenstein:1994jw,ylw,Bowser-Chao:1998yp,wu:2001qu}, the typical absolute
value for $|\xi_t|$ is of the order ${\cal O}(1)$. Taking
$|\xi_t|=1$ and $\delta_t=0$ as an example, the sum of one and two
loop contributions from $h-$scalar exchange reads
 \beqa
 \Delta a_\mu={\alpha \abs{\xi_\mu} \over \sin^2\theta_W}
                        { m^2_\mu \over m^2_W}
                        \left[  {\abs{\xi_\mu}\over 8\pi}
                                 { m^2_\mu \over m^2_h}\ln\left( {m^2_h\over m^2_\mu} \right)\cos 2\delta_\mu
                               - {\alpha\over 3 \pi^2} g\left({m^2_t\over m^2_h}\right) \cos\delta_\mu
                        \right].
\eeqa

%

 It is clear that if $\delta_\mu$ lies in the range $3\pi/4 \leq
 \delta_\mu \leq 5\pi/4$, the one and two loop contributions will be
 constructive.   
  It
 can be seen  that for large values of $\delta_\mu\sim \pi $, the
 contributions from $h-$scalar exchange can reach the experimental lower bound even for  a heavy scalar with $m_h \sim 200$ GeV.  This is quite different from the  existed 2HDM calculations in the literature, where the allowed range  for the scalar boson $A$ must be less than 100 GeV , and the scalar boson $h$ must be much heavier than $A$, so that its negative contributions are negligible.

In summary , the current data on CP asymmetry in \bpk and muon \gmt
can be understood as an implication of new physics beyond the SM. If
the differences between experiment and SM predictions are confirmed
by future high accuracy data, the physics models which have new phases besides the CKM phase are favored. The difference of $\sin 2\b$
extracted from \bpk and global fit can only be explained by the models
with new phase. The large muon anomalous magnetic moment can also be better understood with the inclusion of new phases.


\end{document}